\documentclass[a4paper,pdftex]{jpconf}
\usepackage{graphicx}
\begin{document}
\title{Search for Dark Matter in Upsilon Decays at BABAR Experiment}

\author{Romulus Godang}

\address{Department of Physics \\ 
University of South Alabama \\
Mobile AL 36688, U.S.A.}

\ead{godang@southalabama.edu}

\begin{abstract}
Recent investigations have suggested that the singlet six-quark combination $uuddss$ may be a deeply bound 
state $S$, called Sexaquark. An essentially stable state $S$ is a potentially excellent Dark Matter candidate. 
We present the first search for a stable, doubly strange six-quark state in the decays of 
$\Upsilon(4S) \to \Lambda \bar{\Lambda}$.  Based on a data sample of $\Upsilon(2S)$ and $\Upsilon(3S)$ 
decays collected by the BABAR Experiment we report the most recent results and set stringent limits 
on the existence of such exotic particle.
\end{abstract}

\section{Introduction}

A hexa-quark di-baryon $uuddss$ or $S$ could be a Dark Matter candidate within the Standard 
Model~\cite{Farrar1, Farrar2, Farrar3}.
A large binding energy might make $S$ to be light enough that is stable or long lived. 
The spatial wavefunction of the $S$ is completely symmetric that implies it should be the most tightly 
bound six-quark state of its class~\cite{preskill}. At the same time the color, spin wavefunctions, and 
flavor are totally asymmetric. The $S$ is a spin 0, flavor-singlet, and parity-even boson with Q=0, B=2, and S=-2.  

The $S$ is absolutely stable if its mass, $m_S$, is lighter than $2(m_p + m_e) = 1877.6$ MeV. 
If its mass $m_S < m_p + m_e + m_{\Lambda}$ = 2054.5 MeV, it decays via a doubly-weak interaction and its lifetime could 
be very long. A stable $S$ is allowed by Quantum Chromodynamics (QCD) and would have eluded detection in both accelerator 
and non-accelerator experiments. So far such as bound state $S$ has not been
excluded by hypernuclei decays and direct searches for long-lived neutral state. 
The stable $S$ has not been detected. It is difficult to distinguish the $S$ kinematically from the neutron that 
attributes might explain why this state has escaped detector. The $S$ does not couple to photon, pions, and most of other 
mesons because of its charge neutral and it has a flavor-singlet. The $S$ is probably more compact than the ordinary baryons.

\section{The BABAR Detector}

The BABAR detector was operated at the PEP-II asymmetric-energy storage rings at
the SLAC National Accelerator Laboratory. The data were recorded with the BABAR
detector about 28 fb$^{-1}$ data at $\Upsilon(3S)$ and 14 fb$^{-1}$ data 
at $\Upsilon(2S)$~\cite{Babar_nim3}. Additional samples of an integrated luminosity 
of 428 $fb^{-1}$ collected at $\Upsilon(4S)$ at a center of mass energy of 10.58 GeV
are used to estimate the background.

A detail description of the BABAR detector is presented elsewhere~\cite{Babar_nim2, Babar_nim1}.
The momenta of the charged particles are measured in a tracking system consisting 
of a 5-layer double sided silicon vertex tracker (SVT) and a 40-layer drift
chamber (DCH). The SVT and DCH operate within a 1.5 T solenoid field and have a combined 
solid angle coverage in the center of mass frame of 90.5\%. A detector of internally reflected
Cerenkov radiation (DIRC) is used for charged particle identifications of pions, kaons, 
and protons with likelihood ratios calculated from $dE/dx$ measurements in the SVT and DCH.  
Photons and long-lived neutral hadrons are detected and their energies are measured in 
a CsI(Tl) electromagnetic calorimeter (EMC). For electrons, energy lost due to 
bremsstrahlung is recovered from deposits in the EMC.

\section{Stable Six-Quark State}

We searched the exclusive decay of $\Upsilon(2S,3S) \to S \bar{\Lambda} \bar{\Lambda}$. 
The inclusive six-quark production in the $\Upsilon(2S,3S)$ decays is predicted at the level of
$10^{-7}$ with significant uncertainties. Inclusion of the charged conjugate mode is implied throughout 
this paper. The exclusive decays of $\Upsilon \to S \bar{\Lambda} \bar{\Lambda}$ or $\bar{S} 
\Lambda \Lambda + \pi$ and/or $\gamma$ are ideal discovery channels proposed by Farrar~\cite{Farrar2}. 
No specific prediction of the branching fraction of the decay $\Upsilon(2S,3S) \to S \bar{\Lambda} \bar{\Lambda}$.

The $S$ angular distribution is simulated using an effective Lagrangian based on a constant matrix element
by assuming that angular momentum suppression effects are small~\cite{Farrar4}. The interaction between six-quark states 
and matter is simulated to be similar to that of neutrons. The $\Upsilon(2S,3S, 4S)$ decays events are generated using 
EvtGen~\cite{evtgen}. The detector acceptance and reconstruction efficiency are determined using Monte Carlo (MC) 
simulation based on GEANT4~\cite{geant4}.  

\begin{figure}[h]
\vspace*{-0.4cm}
\includegraphics[width=16cm]{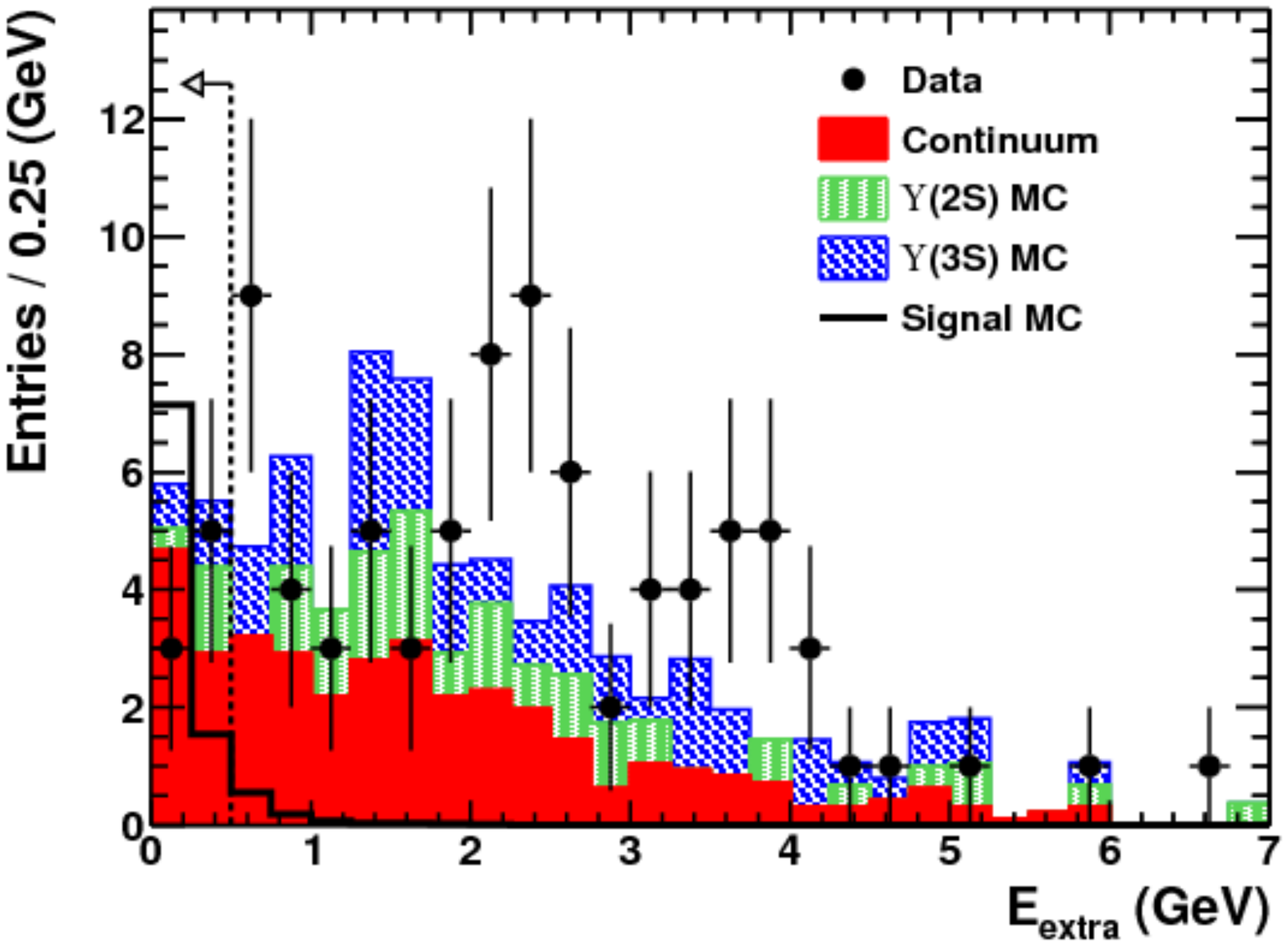}
\vspace*{-3cm}
\caption{\label{label} The distribution of the extra neutral energy ($E_{extra}$), before performing the kinematic fit 
for $\Upsilon(3S)$ and $\Upsilon(2S)$, and various background estimates: continuum (red), $\Upsilon(3S)$ MC (green),
$\Upsilon(2S)$ MC (blue), and signal MC(solid line).}
\label{fig1}
\end{figure}

The events containing at most five tracks and two $\Lambda$ candidates with the same strangeness, consistent
with the topology of the process: $e^+ e^- \to S \bar{\Lambda} \bar{\Lambda}$ final state are selected. 
The events are reconstructed in the $\Lambda \Lambda \to p\pi^- p\pi^-$ final state by requiring 1.10 GeV 
$<m_{p\pi} <$ 1.14 GeV. The additional track not associated with any $\Lambda$ candidate with a distance of 
closest approach (DOCA) from the primary interaction larger than 5 cm is selected. 
The protons and anti-protons are selected by particle identification (PID) algorithms. 
The PID requirement is approximately 95\% efficient for identifying protons and anti-protons and 
removes a large amount of four-pion final state background. 
The total energy clusters in the electromagnetic calorimeter not associated with charged particles, $E_{extra}$, must 
be less than 0.5 GeV. To reduce the contribution of cluster fragments, the distance between the cluster and the proton 
is required to be greater than 40 cm. Figure~\ref{fig1} shows the $E_{extra}$ distribution after applying all selection criteria.  

\begin{figure}[h]
\vspace*{-0.4cm}
\includegraphics[width=16cm]{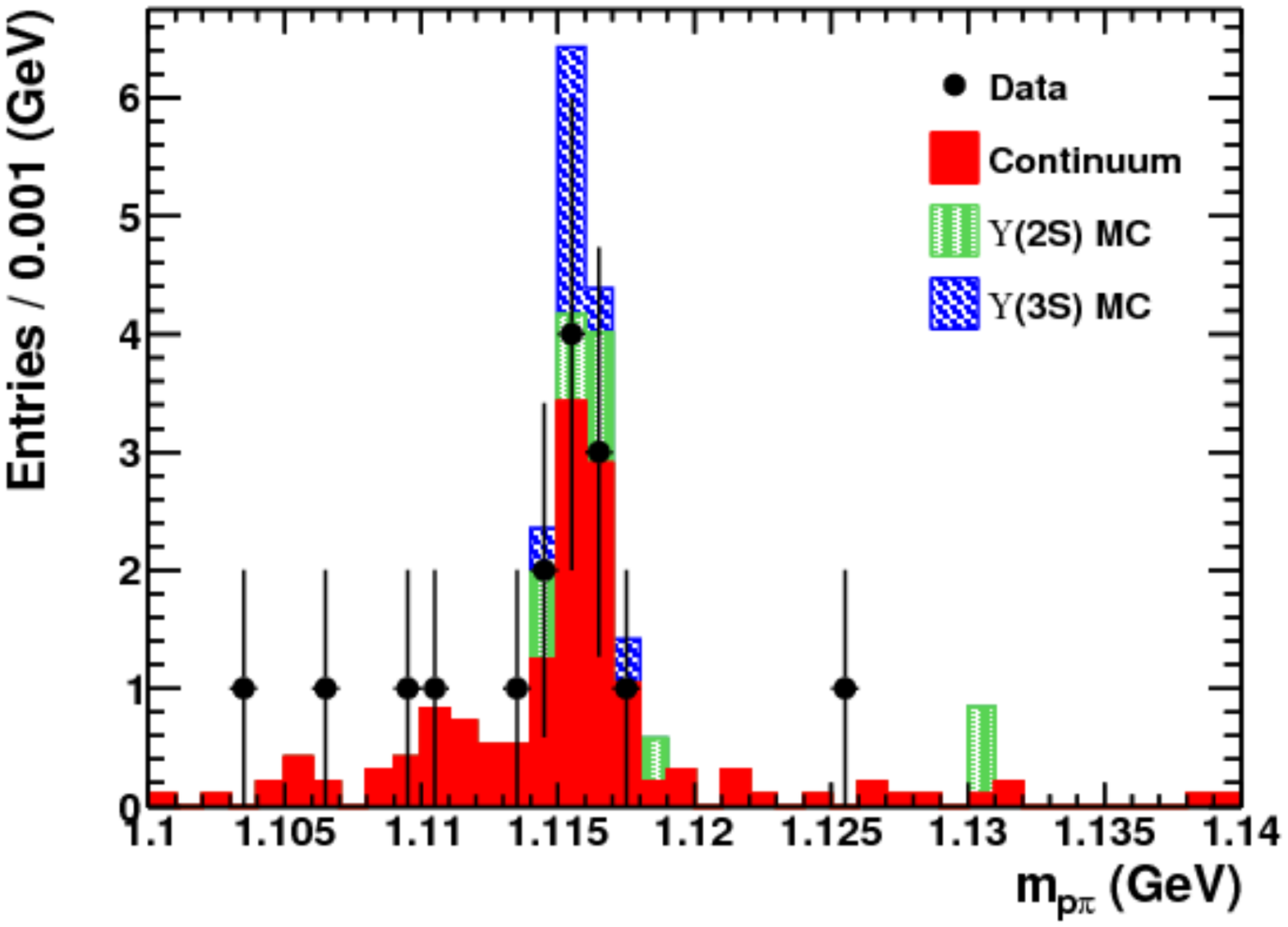}
\vspace*{-3cm}
\caption{\label{label} The distribution of the $p\pi$ invariant mass, $m_{p\pi}$, before performing the kinematic fit 
for $\Upsilon(3S)$ and $\Upsilon(2S)$, and various background estimates: continuum (red), $\Upsilon(3S)$ MC (green),
and $\Upsilon(2S)$ MC (blue).}
\label{fig2}
\end{figure}

To maximize the signal sensitivity the selection procedure is tuned by taking into account the systematic 
uncertainties that are related to $S$ production and the interaction with detector materials. After applying these 
criteria the $p\pi^-$ mass distribution is shown in Fig~\ref{fig2}. A total of eight of $\Upsilon \to S \bar{\Lambda} 
\bar{\Lambda}$ candidates are selected.

\begin{figure}[h]
\vspace*{-0.4cm}
\includegraphics[width=16cm]{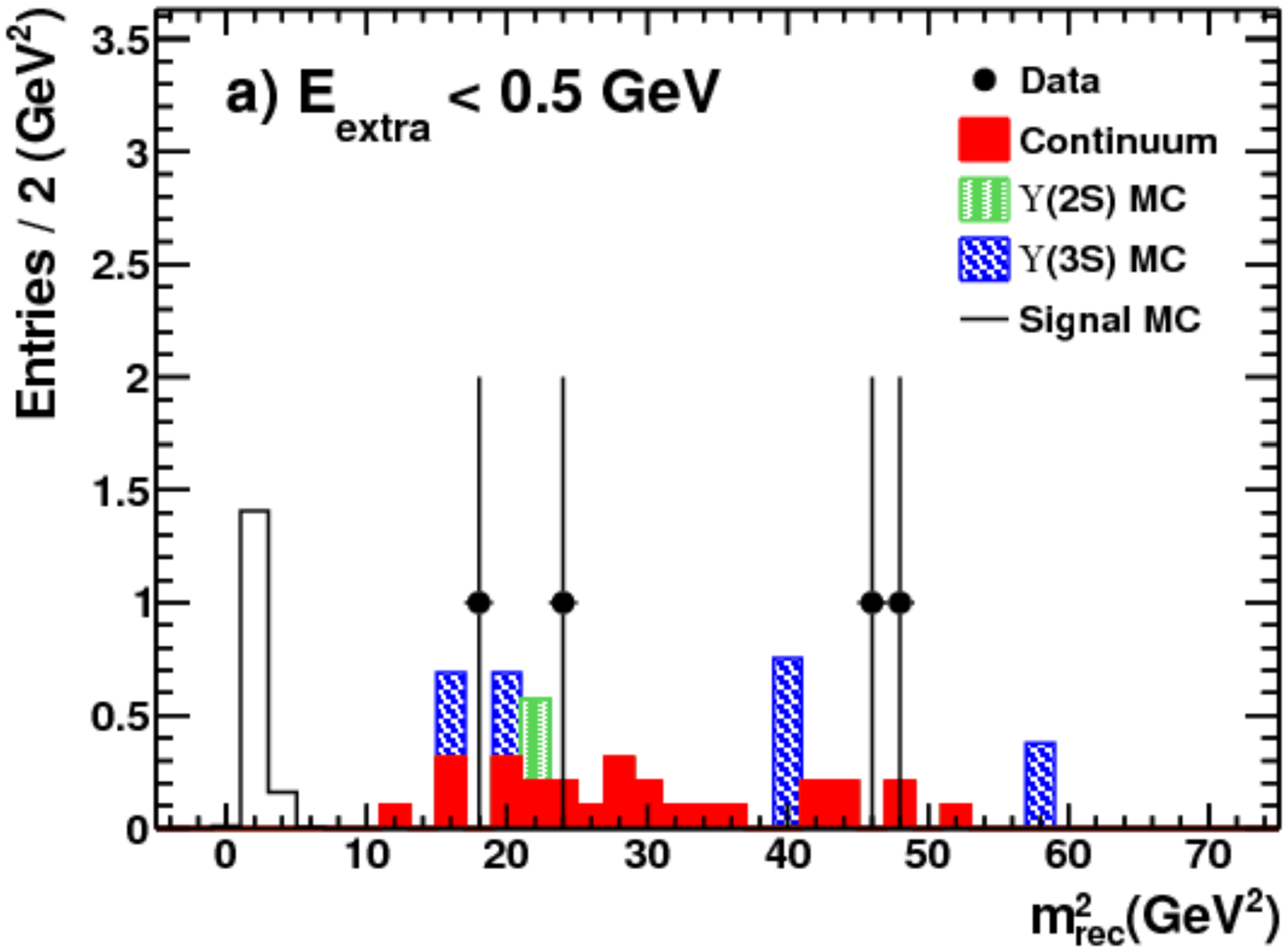}%
\vspace*{-3cm}
\caption{\label{label} The distribution of the recoil mass squared, $m_{rec}^2$, against the $\Lambda \Lambda$ system,
after applying the kinematic fit with various background estimates for the $E_{extra} < 0.5$ GeV signal region.}
\label{fig3}
\end{figure}

We then fit the events by imposing a mass constraint to each $\Lambda$ candidate and requiring a common production 
of the beam interaction point. We select combination with $\chi^2 < 25$, for 8 d.o.f, retaining four signal candidates.
The signal is identified as a peak in the recoil mass squared, $m_{rec}^2$, in the region 0 GeV$^2 < m_{rec}^2 < 5~$ GeV$^2$. 
The recoil mass squared, $m_{rec}^2$ distribution is shown in Fig~\ref{fig3}. 

\begin{figure}[h]
\vspace*{-0.4cm}
\includegraphics[width=16cm]{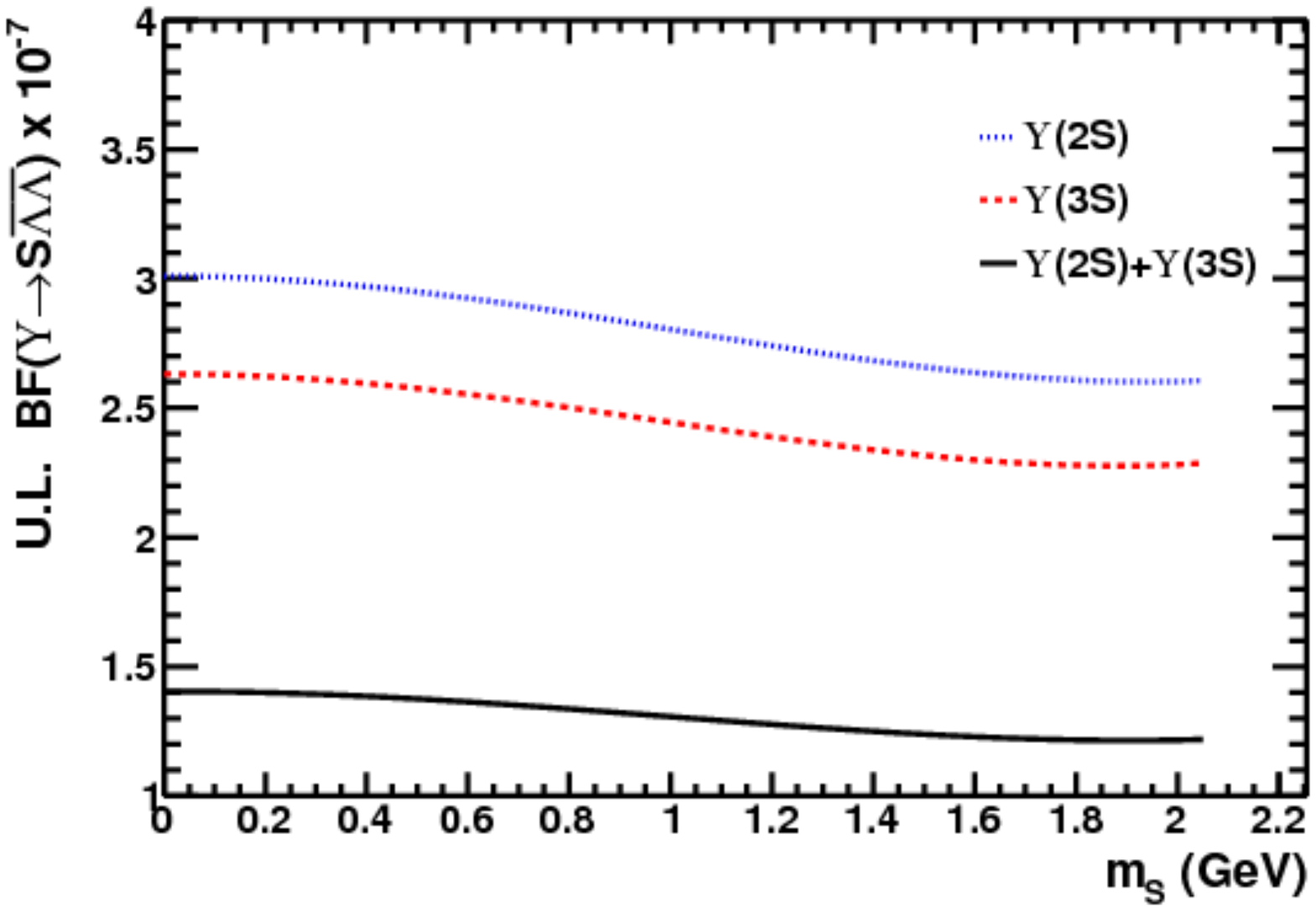}%
\vspace*{-3cm}
\caption{\label{label} The 90\% C.L. upper limits on the $\Upsilon(2S,3S) \to S \bar{\Lambda} \bar{\Lambda}$
branching fractions and the combined of $\Upsilon(2S)$ and $\Upsilon(3S)$.}  
\label{fig4}
\end{figure}

No significant signal is observed. We derive 90\% confidence level (C.L.) upper limits on 
the $\Upsilon(2S,3S) \to S \bar{\Lambda} \bar{\Lambda}$ branching fractions, scanning $S$ masses
in the range 0 GeV $ < m_S < $ 2.05 GeV in steps of 50 MeV as shown in Fig~\ref{fig4}. 
For each mass hypothesis, we evaluate the upper bound from the $m_{rec}^2$ distribution with 
a profile likelihood method~\cite{rolke}.

The main uncertainties on the efficiencies arise from the modeling of the angular distribution
of the $\Upsilon(2S,3S) \to S \bar{\Lambda} \bar{\Lambda}$ to be about $4\%$ and it rises
to $15\%$. The systematic uncertainty due to the limited knowledge of the interactions between the six-quark 
state with matter is estimated from $8\%$ to $10\%$. The systematic uncertainty due to the difference in
$\Lambda$ reconstruction efficiencies between data and MC calculations is $8\%$.  The systematic uncertainty
on the $\Lambda \to p\pi$ branching fraction to be $1.6\%$ and due to the finite MC sample is $1.5\%$.    

\section{Conclusion}

We have performed the first search for a stable six-quark state, $uuddss$ configuration in the $\Upsilon (2S)$ and 
$\Upsilon (3S)$ decays. No signal is observed. We derive 90\% confidence level (C.L.) upper limits on the branching 
fraction of the $\Upsilon(2S,3S) \to S \bar{\Lambda} \bar{\Lambda}$ to be $(1.2-1.4)\times 10^{-7}$~\cite{babar_prl}. 
These results set stringent bounds on the existence of a stable six-quark state. 

\section{Acknowledments}

The author would like to thank the organizers of the $16^{th}$ International Conference on Topics in Astroparticle and Underground
Physics in Toyama, Japan. The supports from the BABAR Collaboration and the University of South Alabama are gratefully acknowledged.

\end{document}